\documentclass[sigplan,nonacm]{acmart}

\usepackage{tikz}
\usepackage{cleveref}
\usepackage{amsmath}

\usepackage{filecontents}
\usepackage{enumitem}

\usepackage{algorithmic}
\usepackage{graphicx}
\usepackage{textcomp}
\usepackage{xcolor}

\definecolor{indigo}{RGB}{51,0,102}
\usepackage{tcolorbox}

\usepackage{hyperref}

\usepackage[framemethod=TikZ]{mdframed}

\usepackage{xspace}
\usepackage{scalerel}
\usepackage{ctable}

\usepackage{pifont}% http://ctan.org/pkg/pifont

\usepackage{url}
\usepackage{mathtools}

\usepackage{tabularray}

\usepackage{multirow}
\usepackage{array}
\usepackage[normalem]{ulem}

\newcommand{\criteria}[1]{{\large \textcircled{\small #1}}}

\newcommand{\newtext}[1]{#1}

\AtBeginDocument{%
  }

\setcopyright{acmlicensed}
\copyrightyear{2018}
\acmYear{2018}
\acmDOI{XXXXXXX.XXXXXXX}

\acmConference[ASPLOS '25]{}{June 03--05,
  2018}{Woodstock, NY}
\acmISBN{978-1-4503-XXXX-X/18/06}

\begin{document}
\title{Pedagogically Motivated and Composable Open-Source RISC-V Processors for Computer Science Education}

 \author{Ian McDougall}
 \affiliation{%
   \institution{University of Wisconsin-Madison}
   \country{USA}
 }

 \author{Harish Batchu}
 \affiliation{%
  \institution{University of Wisconsin-Madison}
  \country{USA}
 }

 \author{Michael Davies}
 \affiliation{%
   \institution{NVIDIA Research}
   \country{USA}
 }

 \author{Karthikeyan Sankaralingam}
 \affiliation{%
   \institution{University of Wisconsin-Madison}
   \country{USA}
 }

%\author{Authors}
%\email{Emails}
%\affiliation{
%  \institution{Affiliations}
%  \city{City}
%  \state{State}
%  \country{Country}
%}


\begin{abstract}
While most instruction set architectures (ISAs) are only available to use through the purchase of a restrictive commercial license, the RISC-V ISA presents a free and open-source alternative. Due to this availability, many free and open-source implementations have been developed and can be accessed on platforms such as GitHub. If an open source, easy-to-use, and robust RISC-V implementation could be obtained, it could be easily adapted for pedagogical and amateur use. In this work we accomplish \newtext{three} goals in relation to this outlook. First, we propose a set of criteria for evaluating the components of a RISC-V implementation's \textbf{ecosystem} from a pedagogical perspective. Second, we analyze a number of existing open-source RISC-V implementations to determine how many of the criteria they fulfill. \newtext{We then develop a comprehensive solution that meets all of these criterion and is released open-source for other instructors to use. The framework is developed in a composable way that it's different components can be disaggregated per individual course needs. Finally, we also report on a limited study of student feedback.}
%Through the elaboration of the ecosystem model and the results of our analysis, this paper aims to serve as a pedagogical roadmap for instructors to determine if a particular RISC-V implementation is useful for their specific needs.
\end{abstract}

\maketitle
\pagestyle{plain} % should come right after \maketitle

\section{Introduction}
%The Instruction Set Architecture (ISA) is perhaps the most important component in the design of a computer. It constitutes the interface between software, and the hardware it runs on. This enables software to be compatible across successive hardware generations, decoupling software development from hardware.

%The most popular ISAs found today are commercially licensed (E.g. x86 and ARM), the cost and scope of which ranges from incredibly restrictive (x86) to ``more accessible" but a legal minefield (ARM Ltd. has been known to sue institutions which use their ISA for educational purposes if they do not pay the license fee, as well as disallows individuals from making their own implementations public). In an effort to sidestep both legal and financial challenges many attempts to build a ``free and open-source'' ISA have been presented over the last few decades as a valuable alternative -- perhaps the most popular of these being RISC-V~\cite{rv-foundation}.

%RISC-V originally was developed out of a research group at the University of California-Berkeley and has grown to be used in many research groups across the world, by amateurs and hobbyists, and even in many commercial applications~\cite{wd-riscv,ditzel_accelerating_2022}, both established companies and startups. Many of these applications are freely accessible on code repositories such as GitHub. The development group behind RISC-V provides documentation (The ``RISC-V Exchange'') which contains an index of a large number of public RISC-V microarchitecture projects ~\cite{riscv-exchange}. 

% shortened paragraph
The Instruction Set Architecture (ISA) is the interface between software and hardware. It is the most important component in the design of a computer, as it enables software to be compatible across successive hardware generations. The most popular ISAs today are commercially licensed (e.g. ARM and x86), but there have been many attempts to build a free and open-source ISA as an alternative. One of the most popular of these is RISC-V, which was originally developed out of a research group at the University of California-Berkeley~\cite{rv-foundation}. RISC-V is used in many research groups, by amateurs and hobbyists, and even in many commercial applications~\cite{wd-riscv,ditzel_accelerating_2022}. The development group behind RISC-V provides documentation and an index of a large number of public RISC-V microarchitecture projects~\cite{riscv-exchange}.

An \textbf{ecosystem} comprising of the software stack, tooling, and compiler toolchain must accompany an ISA's implementation in order for it to be useful in a pedagogical environment -- either built from scratch by an educator, or reused from open-source projects. 
Introductory architecture courses typically involve exposure to an ISA, as well as implementation, and this necessitates having software (ecosystem) components such as a compiler, test programs, and simulation testbench for students to use in evaluating the correctness of their design. Section~\ref{sec:criteria} elaborates more on the components and necessity of this ecosystem. This paper investigates the role of such an eco-system in the context of introductory digital systems courses in the ACM curriculum \cite{acm_computing_curricula_task_force_computer_2013}.

Many university courses use self-developed ISAs or pre-existing free and open-source ISAs with a mix of ground-up or pre-made ecosystems being employed, including the following examples: CS/ECE 552 at the University of Wisconsin-Madison~\cite{csece552wisc}, CE4304 and CS2340 at University of Texas at Dallas~\cite{ce4304ut,cs2340ut}, CSE469 at the University of Washington~\cite{cse469uw}, ECE 411 at the University of Illinois Urbana-Champaign~\cite{ece411uiuc}, and ECE-M116C/CS-M151B at University of California, Los Angeles~\cite{eceucla}. With the recent renewed interest in hardware and semiconductors, many are being revised to reflect the modern role of architecture in industry. RISC-V is an attractive path forward to "modernize" these offerings.

With ecosystem level components such as a compiler being costly to build from scratch, there is incentive to reuse existing components as much as possible\footnote{Plagiarism is certainly a concern with readily available reference implementations when using a publicly available ecosystem, a ground-up ecosystem would not solve this issue since other reference designs for open ISAs exist and can be adapted to fit. Former students also often make their projects available publicly for course-specific ground-up ISAs.}. However, we observe there is a lack of in-depth information on the \textbf{ecosystem} surrounding popular RISC-V implementations. 
% Our solution
In this work, we intend to bridge this gap and construct a \textit{pedagogical roadmap} for introductory computer architecture education. We do this by developing a set of criteria for assessing the features and ecosystem of RISC-V implementations and then conduct a wide study on the basis of these criteria of many open-source RISC-V designs, including detailed analysis of core microarchitecture, building and running host system software and verification infrastructure, and running FPGA synthesis. To the best of our knowledge, this is the first and most comprehensive such analysis of available RISC-V implementations. We then develop a complete system stack called WISCV\footnote{For anonymized review it expands to Well Intentioned \& Systematic Computer in Verilog} of all the components that can be deployed in a turn-key fashion, and disaggregated per instructor needs. This includes multiple reference implementations of a working RISCV (pipelined, single cycle, and with cache) core, skeleton versions that students can start with (as part of their implementation), verification infrastructure, test generators, and FPGA mapping infrastructure on an inexpensive \$159 Arty A7 board. Finally, we also report on a limited study of student feedback from using it in a course. \textbf{Our infrastructure is available at \url{https://anonymous.4open.science/r/wiscv-86E1}}. It will be hosted and maintained publicly on github post review.

The rest of this paper is organized as follows. Section~\ref{sec:criteria} explains the ecosystem and develops and explains a useful set of objective criteria for the purpose of evaluating an open-source RISC-V implementation's ecosystem. It also describes the scope and methodology we employ to evaluate a large number of popular open-source RISC-V designs. Section~\ref{sec:results} presents the results of this survey and analysis, and discusses the current state of open-source RISC-V microarchitecture development from a computer science education perspective and roadmap for various different use cases. The repositories of each microarchitecture design can be found online~\cite{rocket,pulpino,ibex,pulpissimo,mriscv,steel,biriscv,darkriscv,serv,cv32,picorv32,cva6}\footnote{It is important to note that many of these implementations are in active development, so their current state may not reflect the findings of our survey}. Further, the authors of this paper are not affiliated with any of these projects. Section~\ref{sec:wiscv} describes our WISCV implementation, relating it to our previously developed ecosystem. In this section we also present student feedback regarding our WISCV implementation.

\section{The RISC-V Ecosystem}\label{sec:criteria}
In this section, we first describe the scope of the ecosystem which surrounds a RISC-V implementation, and the imputed needs on this ecosystem from a pedagogical perspective. 

\subsection{Defining the Ecosystem}
ACM curriculum \cite{acm_computing_curricula_task_force_computer_2013} recommends an introductory digital systems course cover a few specific topics, including the design of combinational and sequential logic, register transfer logic (RTL) notation in Verilog (or VHDL), computer-aided design, and an overview of architecture, its history, and the many layers of abstraction in a computer. From this, it is possible to abstract the primary goals of computer architecture courses as: (1) learning ISA basics, (2) learning microarchitecture basics, and (3) implementing a basic computer processor, very often mapping it to an FPGA. It is worth noting here that mapping a computer processor design to a FPGA in the context of such courses is an important step which introduces the complexities of actual hardware synthesis rather than the idealizations of software simulations, and serves as a capstone experience for their processor design project. Furthermore, educational psychology has shown that active learning by building artifacts helps in better learning outcomes~\cite{lester_pedagogical_1997}. While inconceivably complex, expensive and time-consuming a decade or so ago, modern FPGA hardware and frameworks, have brought this (potentially) down to push-button capability making this a possibility for undertaking in undergraduate courses. 

\begin{figure*}[htp]
    \centering
    \includegraphics[width=\textwidth]{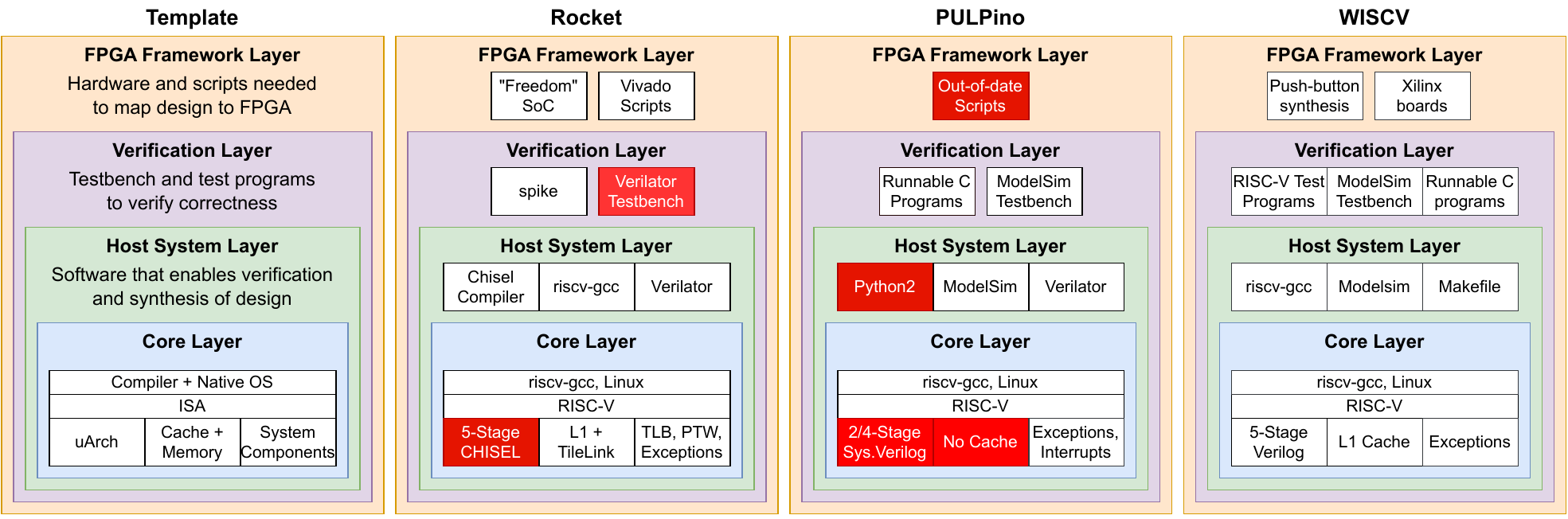}
    \caption{Ecosystem layers, two examples, and WISCV which satisfies all pedagogical goals. Red boxes indicate deficiencies.}
    \label{fig:ecosystem}
\end{figure*}

The ecosystem comprises of four layers as enumerated below and described in depth in the successive subsections, and within each we have a set of criteria to be evaluated. Figure~\ref{fig:ecosystem} shows this ecosystem with example implementations showing which portions are covered in which implementations (red denotes deficient).
\begin{enumerate}
    \item \textbf{Core Layer} -- Encompasses the ISA, core pipeline design, memory components and system components, and compiler and operating system\footnote{These final two are treated in their own separate courses and so we do not focus on them in this paper.}.
    \item \textbf{Host System Layer} -- Encompasses all of the support software which must run on the host system in order to run simulations, aid debugging and verification, and synthesis.
    \item \textbf{Verification Layer} -- Encompasses test programs and testing infrastructure (including a simulator testbench) to validate the correctness of the core.
    \item \textbf{FPGA framework layer} -- Encompasses wrapper RTL, synthesis scripts and related software to "map" a design and run it on an FPGA.
\end{enumerate}
% 1) the core layer, which consists of the actual design of the processor, including: (i) microarchitecture, (ii) ISA, , and (iii) DUT system, (2) the host system layer, which includes the software necessary to compile the system on a host machine, (3) the verification layer, which includes testbenches and test programs used to verify the design's correctness, and (4) the FPGA framework layer, which includes the software and hardware specifications necessary to map the design to a FPGA. While for this sort of ecosystem the ISA can only be fulfilled by RISC-V, and so is a trivial requirement, the other categories each contain nuances which require further elaboration. The following sections will provide such an elaboration for each element, along with a justification for their inclusion within the proposed ecosystem model.

\subsection{Defining the Criteria}
The following criteria are used to assess an implementation's ecosystem. We organize them based on the layers previously introduced.\\
\\
\noindent \textbf{Core Layer.}
\criteria{1} Basic 5-Stage Pipeline. 
\criteria{2} Branch Prediction. 
\criteria{3} Caching / Variable latency memory support.
\criteria{4} Written in Verilog.
\\
\noindent \textbf{Host System Layer.} 
\criteria{5} Compatible with modern Linux systems
\\
\noindent \textbf{Verification Layer.}
\criteria{6} Working testbench.
\criteria{7} Working test Programs.
\\
\noindent \textbf{FPGA Layer.}
\criteria{8} SoC wrapper for FPGA synthesis.
\criteria{9} Out-of-box mapping to inexpensive FPGA.
\criteria{10} Functional FPGA Framework.
\\
\noindent \textbf{Research Layer.}
\criteria{11} Extendable pipeline depth.
\criteria{12} High Frequency.
\criteria{13} Extendable functionality.

% \begin{itemize}
%     \item Core Layer
%     \begin{itemize}
%         \item Basic 5-stage pipeline (1)
%         \item Branch Prediction (2)
%         \item RTL is Verilog (put this in system layer?) (3)
%     \end{itemize}
%     \item Host System Layer
%     \begin{itemize}
%         \item Compatible with Modern Linux/libc (4)
%     \end{itemize}
%     \item Verification Layer
%     \begin{itemize}
%         \item Working testbench (5)
%         \item Working test programs (6)
%     \end{itemize}
%     \item FPGA Framework Layer
%     \begin{itemize}
%         \item Full SoC Support (7)
%         \item Out of the Box Mapping to Cheap FPGA Chips/Cards (8)
%         \item Easy FPGA Synthesis (9)
%     \end{itemize}
% \end{itemize}

% Additional Requirements for Amateur Research Use (remove?)
% \begin{enumerate}
%     \item Extensibility for deeper pipeline (10)
%     \item High Frequency (11)
%     \item Extendibility (12)
% \end{enumerate}

\subsection{The Core Layer}
The core layer includes the ISA, the microarchitecture implementation, cache and memory system, and system-level components. We only focus on RISC-V since it is one of the most popular open ISAs and in addition, observe that specific choice of ISA makes little difference in real-world applications~\cite{hpca2013:power-struggles}. The microarchitecture encompasses the core pipeline structure, the presence of branch prediction, the modules by which the instructions are decoded and interact with memory systems, etc. From a pedagogical perspective, we identify that \criteria{1} a basic 5-stage pipeline, \criteria{2} a branch predictor, and \criteria{3} a caching system, are the three key features of the microarchitecture needed to sufficiently cover basic processor microarchitecture in an introductory setting. This set of features typically fits nicely into a standard semester-length course and handily covers most of the ACM curriculum by exposing students to RTL design entry, debugging, logic analysis, and architecture.
% This is because most course textbooks present a 5-stage pipeline as their main case study, and branch-prediction and caching as intermediate level features \cite{comp-org-textbook}.

We find that choice of hardware description language (HDL) is also important since it impacts the kinds of tools and verification methodologies that can be used in a real-world environment. While there are instances of custom HDLs being used in the classroom, such as Chisel\footnote{Chisel adds hardware construction primitives to the Scala programming language.} \cite{chisel}, we take the view that an implementation in plain Verilog \criteria{4} is best suited for introductory course material. This is because Verilog is the most commonly used HDL in production environments, has the best tooling and simulators, and is most amenable to industry-standard verification methodologies\footnote{We acknowledge SystemVerilog also fills these requirements but believe Verilog is a better choice for an introduction to RTL design entry since it is much simpler, and is a strict subset of SystemVerilog.} \cite{1620780}.

\subsection{The Host System Layer}
The Host System Layer encompasses all of the support software needed to simulate and verify the design on some host machine. The existence and quality of this tooling is particularly important in the classroom since it is the methodology by which students will experiment with their own designs and verify correctness. Concretely, we identify that the host software, simulator, and tooling must \criteria{5} be able to run on modern Linux platforms since this is the most commonly found on campus computing resources. This criteria actually proves to be a surprising challenge for some designs, and is discussed further in Section \ref{sec:results}.

% For this implementation and verification to take place, however, there needs to exist some simulation, hardware synthesis, and verification software, all of which interacts with the system of the host machines found in the classroom. Therefore, for this layer, our sole criterion is that the software included in the project be compatible with up-to-date Linux and libc stacks. This is important because instructors shouldn't be expected to run out-of-date software stacks to support software used in a particular course, especially if they are instructing multiple courses during a given timeframe.

\subsection{The Verification Layer}
The verification layer includes test programs -- compiled binaries which can execute on the design and are aimed at exposing functional bugs -- and the testbench -- a combination of software and behavioral RTL to facilitate executing these test programs and extracting debugging information. In a classroom setting this is important for the students as well as educators. For students, the testbench and test programs facilitate bug discovery and design iteration. For educators this is the primary path to evaluating the correctness of students' work for the purpose of grading. We identify two criteria are needed here -- \criteria{6} a testbench exists and works on modern computer equipment, and \criteria{7} existing test programs are provided and work on the design out-of-box. We also need a reference design (like RISC-V instruction emulator) and a trace generation infrastructure that creates semantically equivalent traces from the RTL and the reference design, which is then compared to deem the RTL implementation correct. This method of using checkers is industry standard methodology~\cite{uvm}.

% The verification layer consists of all provided testbenches and test programs which will allow for verification of the given design. For our purposes, a ‘testbench’ is a piece of software which tests the emulated behavioral model of the design, while a ‘test program’ is a binary which can run on actual synthesized hardware, but which could also be tested via emulation. The software we were specifically looking for, i.e. testbenches and test programs, were considered because of their usefulness to educators. If, for instance, students were tasked with creating a modified version of a particular design, testbenches and test programs could be used by the instructors to verify the functionality of that design. Both are useful, as they provide a means to test students’ designs both in emulation and synthesis. Therefore, for this layer, our two criteria were the presence of working testbenches and the presence of functional test programs.

\subsection{The FPGA Framework Layer}
The FPGA layer contains additional RTL, scripts, and metadata to synthesize a design to run on an FPGA development board. The additional RTL needed contains glue logic which interfaces the design with the particular memory system of a development board (E.g. DRAM controller with AXI interface), as well as facilitates communication with a host machine for the purposes of debugging (E.g. JTAG, UART, etc.). The design, along with this additional logic, constitutes a full system-on-chip (SoC) and this ``SoC-wrapper'' \criteria{8} logic is a necessary component. In addition to this, the included scripts and meta-information for FPGA mapping must be suited for an inexpensive FPGA board \criteria{9}. We impose this requirement since universities typically must buy many development boards to allow all students the ability to work on their projects, so being tied to an expensive FPGA part would be prohibitively expensive for a university to employ at scale in the classroom. Finally, we identify that included synthesis scripts and meta-information must work out-of-box \criteria{10}\footnote{Unlike software-based architectures like GPU chips/cards, FPGA cards are less straight-forward from a system and usage perspective of getting things to run, usually involving many person-months for bringing up a new design. In short, while a simple {\tt make} on the downloaded SDK would get one going on running a CUDA program on a GPU, or C program on a Raspberry Pi for example, the process is far more complex and time consuming for an FPGA.}.

\begin{figure*}[t]
    \centering
    \includegraphics[width=1\textwidth]{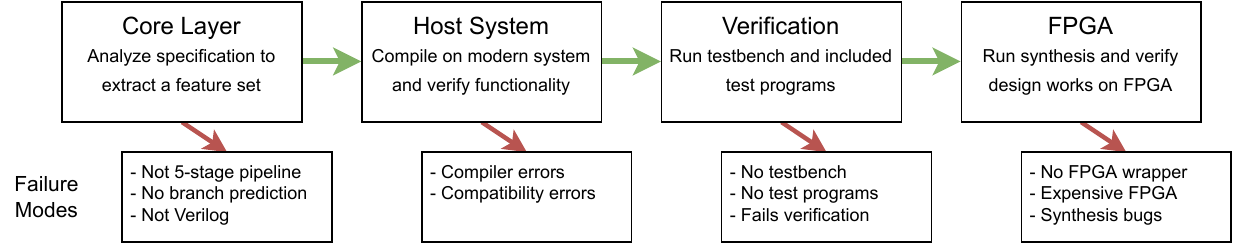}
    \caption{Research Methodology}
    \label{fig:methodology}
\end{figure*}

%\subsection{Research Criteria} In academic research, it is often desired to make modifications to an existing implementation's pipeline structure for the purpose of rapidly testing new features. 
Somewhat orthogonal to teaching, for research purposes additional constraints are:
an easily expandable pipeline \criteria{11}, design for high frequency \criteria{12} in order to give an apples-to-apples comparison with production state-of-art microprocessors, and easy integration with other modules like complex memory hierarchies \criteria{13}.

% For this, we needed to ensure that implementing a deeper pipeline was possible, as this was not always the case, as shall be seen in sections 5.1 and 5.2. High frequency was another important criterion, in order to make sure that these designs could be legitimately useful for amateurs and researchers, and not be entirely restricted for use as classroom projects. Furthermore, extendibility was added as a criterion in order to see more generally if the microarchitecture could be expanded. For example, if a microarchitecture only consisted of a CPU core and did not include caches, a memory interface or I/O, we tried to determine if those features could be added to the design.

\subsection{Research Methodology}\label{sec:methodology}

In order to complete our survey, we first identified a list of RISC-V implementations which we believe were good candidates to study. We started with two of the most well-known free-and-open RISC-V SoC designs: Rocket (and its related SoCs developed by the CHIPS Alliance), and the PULP Platform (developed by researchers at ETH Zürich and the University of Bologna). From this, we made use of the RISC-V Exchange to identify less well-known designs. For a smaller-scale design to be considered, it had to be written in Verilog (See Section \ref{sec:criteria}), and it had to be under a FOSS license (e.g. GPL, MIT, BSD, etc.). This information (HDL and license type) is provided by the RISC-V Exchange.

For each design in our compiled list, we performed the following steps: 1) examined their specification and code repository to determine relevant features of the core microarchitecture \criteria{1}-\criteria{4}, 2) attempted to compile the design on an up-to-date Linux machine \criteria{5}, 3) attempted to run any testbench or test programs present \criteria{6}-\criteria{7}, and 4) attempted to map design to one of two popular and inexpensive FPGA boards we had on hand \criteria{8}-\criteria{10}: a ZED Board or an Arty A7-35T\footnote{If a project had a specific dependence and compatibility with a different FPGA board, we were unable to confirm  if it really did run on that specific part. We observe here that such FPGA board/card specific dependence is poor design philosophy and theoretically should not arise}. As a part of this phase of analysis, we also recorded all errors we encountered. Figure~\ref{fig:methodology} visualizes this process.

% When a RISC-V design passed these initial checks, we then analyzed its code repository and associated documentation to check for more criteria. In most cases, this analysis would enable us to determine if a given design (a) lacked a basic 5-stage pipeline, (b) lacked test programs or a testbench, (c) lacked support for modern Linux/libc versions, (d) lacked support for relatively cheap FPGA boards, and (e) the functionality of the design (e.g. the presence of branch prediction or caches).

% After completing this analysis, we would then download the code associated with a given microarchitecture design and attempt to complete tasks (iii)-(v) as mentioned above. 
% For FPGA mapping (step 4) 

%For the research criteria \criteria{11}-\criteria{13}, we did not perform as detailed an analysis, as we considered those criteria secondary in comparison to the first nine. For \criteria{12} (high frequency) we took information from official documentation, where possible. If the microarchitecture successfully built on our FPGA board, we then used Vivado to analyze the actual processor speed of the design. For \criteria{11} and \criteria{13}, we hypothesized about the possibility of further extensions to the pipeline and the design’s extendibility based on our observations of the design and from information gleaned from official documentation. In many cases, we were unable to come to a definite conclusion without further study. 

\begin{table*}
\centering
  \caption{Presence of criteria 1-9 as listed in 2.1.}
  \label{tab:1}
  \begin{tabular}{|c|c|c|c|c|c|c|c|c|c|c|c|c|c|c|c|}
    \hline
    Implementation & Creation Date & Last Modified Date & \criteria{1} & \criteria{2} & \criteria{3} & \criteria{4} & \criteria{5} & \criteria{6} & \criteria{7} & \criteria{8} & \criteria{9} & \criteria{10} & \criteria{11} & \criteria{12} & \criteria{13}\\
    \hline
    Rocket & October 2011 & March 2021$^\dagger$ & Y & Y & Y & N & Y & Y & Y & Y & Y & Y & N & ? & ?\\
    PULPino & August 2015 & May 2019 & N & N & N & N & N & N & Y & Y & Y & Y & ? & ? & ?\\
    PULPissimo & February 2018 & November 2022 & N & Y & Y & N & Y & Y & N & Y & N & N & Y & Y & N\\
    Ibex & April 2015 & August 2023 & N & Y & N & N & Y & Y & N & Y & Y & Y & Y & Y & Y\\
    mriscv & September 2016 & February 2018 & N & N & Y & Y & N & Y & N & Y & N & N & ? & ? & ?\\
    Starsea-riscv & September 2020 & ??$^\alpha$ & N & N & N & Y & Y & Y & Y & Y & N & N & Y & ? & ?\\
    Steel-core & May 2020 & August 2023 & N & N & N & Y & Y & Y & Y & Y & Y & Y & Y & ? & ?\\
    biriscv & February 2020 & September 2021 & N & Y & Y & Y & Y & Y & N & Y & Y & N & N & Y & N\\
    darkriscv & August 2018 & August 2023 & N & N & Y & Y & N & N & N & Y & Y & N & Y & Y & ?\\
    serv & October 2018 & July 2023 & N & N & N & Y & Y & Y & N & Y & Y & Y & Y & N & Y\\
    CV32E40P & April 2015 & August 2023 & N & Y & N & N & Y & Y & N & N & N & N & Y & Y & N\\
    picorv32 & June 2015 & January 2022 & N & N & Y & Y & Y & Y & Y & Y & Y & Y & N & Y & ?\\
    cva6 & October 2017 & August 2023 & N & Y & Y & N & N & Y & N & Y & N & Y & ? & N & ?\\
  \hline
\end{tabular}

$^\dagger$This is the last modified date for `Freedom', the Rocket Chip SoC used in this paper \\ 
$^\alpha$No longer publicly available on Github as of August 2023
\end{table*}

\section{Analysis of Implementations}\label{sec:results}
This section presents and explains the results of our analysis, which can be found compactly in Table \ref{tab:1}, covering Rocket and PULP in depth, and summarizing the rest. 
\subsection{Rocket}
The Rocket Chip was an incredibly easy implementation to get up and running, and provided a number of microarchitectural features which perfectly fulfilled our criteria, such as a five-stage pipeline, branch prediction, and caching. Despite this, however, it ultimately did not fulfill our proposed RISC-V ecosystem. The primary issue with the Rocket Chip was that it was entirely written in Chisel, with only a small number of simulated components being written in Verilog. As Chisel is not an industry standard RTL, we felt that it would be inappropriate for the Rocket Chip to be used pedagogically. While researchers and amateurs may desire to learn Chisel in order to utilize the advantages presented by the Rocket-Chip family of SoCs, it seems more likely that they will rely on industry standard RTLs. Additionally, as outlined in ~\cite{10.1145/3470496.3533040}, Chisel is impractical for commercial purposes due to various limitations, further limiting its desirability from a pedagogical standpoint. While it is true that a part of the Rocket-Chip workflow is the automatic translation of Chisel into Verilog, the code produced was not human readable. Furthermore, parts of the SoC design, such as the memory interface blocks, could not be directly described by Verilog, meaning that a ‘transcription’ of the SoC design that would fully meet our criteria seems to be impossible, at least if the integrity of the current Rocket-Chip design is to be preserved. \newtext{Rocket's test generator scripts and tests provided are also not very extensive.}

\textit{Summary Finding: Rocket implements a complete ecosystem, but lacks the modularity to allow the higher layers of the ecosystem to be used in tandem with a custom Verilog core design. Verification is also deficient.}

\subsection{The PULP Platform}
Compared to the Rocket-Chip, the PULP Platform had a number of obvious advantages and disadvantages. The two PULP SoC designs we analyzed were the PULPino and the PULPissimo. Both of these designs were written with SystemVerilog, which, while not meeting our strict criteria, was at least an industry standard RTL. However, PULPino either used the RISCY CPU design, which has a four-stage pipeline (lacking a nicely encapsulated memory access stage which is important from a pedagogical perspective), or the zero-riscy CPU, which has a two-stage pipeline. Similarly, PULPissimo can use either RISCY or Ibex, an updated version of zero-riscy. It does seem, however, that there is nothing stopping a deeper pipeline being implemented in either of these cores. The more major issues arose when we attempted to build these systems locally. 

One barrier, at least to amateurs, is that the PULP Platform requires ModelSim and VCS for its simulation systems, which has restrictive licensing costs. PULPino also suffers from issues due to the fact that it uses out-of-date versions of software and lacks support, as the PULP team is now actively developing PULPissimo instead. As a result, more current versions of ModelSim do not seem to support the version of GCC needed by PULPino. Similarly, FPGA synthesis for PULPino only seems to work with out-of-date versions of Vivado \footnote{tested on 2015.1}. Furthermore, in order to generate the BOOT.BIN file, an out-of-date version of the Xilinx SDK is needed, which isn’t supported by current versions of Linux (Ubuntu). On the other hand, PULPissimo~\cite{pulpissimo} suffers from the following issues: inability to generate bitstreams even for the supported FPGAs, very large memory need from FPGA ($>8GB$), missing test programs, and use of an unfamiliar dependency manager, Bender.

%PULPissimo, although actively supported, also suffers from some issues. We were unable to generate bitstreams for a number of the FPGA boards it supported with the up-to-date version of Vivado, and the only cheap board that seemed to support synthesis was not a general-purpose FPGA (that board being the Nexys Video). When bitstream generation did work, it seemed relatively resource intensive, needing greater than 8GB of RAM, which is more than what other RISC-V microarchitectures we built required.  Another major issue is that the official test programs, which are supposed to be downloaded via script, seem to be locked behind a PULP team member’s Github account, meaning they are not publicly accessible. Lastly, PULPissimo is implementing its own dependency management tool, Bender, which introduces additional difficulties if used, since one would need to implement a modified version of the CPU RTL with Bender as part of the workflow. This adds an additional layer of knowledge required to correctly use this system. As of the time of this paper it was not the default setting, but that is a goal which the PULP team has stated \cite{pulpissimo}. 

\textit{Summary Finding: the Host System Layer, Verification Layer, and FPGA layer were riddled with inconsistencies and need substantial overhaul and rewrite to be usable.}

\subsection{Other RISC-V Designs}
After realizing that our criteria were not met by either Rocket or PULP, we analyzed a number of lesser-known RISC-V open-source microarchitectures which could be found on RISC-V’s official ‘RISC-V Exchange’. We specifically analyzed cores and SoCs which used permissive free software licenses (e.g. MIT and BSD). The designs we found had numerous strengths and weaknesses, but none fully satisfied our criteria laid out in Section \ref{sec:criteria}. It is possible that other RISC-V cores exist which more fully meet our criteria, but if they do, they neither appear on the RISC-V Exchange nor use permissive free software licenses.

Of these nine designs, about half were able to be built on modern Linux machines using easily obtainable software. The other half either required simulation and synthesis software with restrictive licensing costs (e.g. darkriscv), toolchains which didn’t support up-to-date Linux versions (e.g. mriscv), or out-of-date synthesis software (e.g. cva6). Beyond this we also encountered software errors with many designs at varying layers of the ecosystem, including: failing to make particular versions of included RISC-V compiler toolchains, test programs being unable to be simulated by Verilator or simply not compiling, and the synthesis script failing to map the implementation to a supported FPGA after execution. Most implementations also included testbenches and testF programs, but those that had limited simulation support failed to generate the tests. More than half of the designs were able to be synthesized onto relatively cheap FPGAs. It is also worth noting that some design repositories stated that they used an out-of-date Vivado version, but still worked with up-to-date versions. 

Because most of these designs were either designed by individuals or small groups there were very little troubleshooting resources available online. Many designs were also no longer actively maintained by their developers, which was the main reason many of them relied on out-of-date versions of software.    
The majority of the designs we examined used Verilog as the HDL, while three used SystemVerilog. None of these designs used the five-stage pipelined model, and many gave no particular reason why they chose the CPU model they implemented. Two designs (mriscv and picorv32) didn’t seem to use any noticeable pipelining, while another (serv) was designed to be a bit-serial processor. Some designs also didn’t include caches, although all except serv, mriscv, and picorv32 (due to the inherent difference in design philosophy) probably could be modified to become a standard five-stage pipelined CPU design with some effort. It should also be mentioned that most of these designs also touted high frequencies (50MHz-450MHz), but most were impossible to test as they didn’t support the FGPAs we had available to us or were unable to be synthesized.

\textit{Summary Finding: All other implementations only satisfied various subsets of criteria, with only about half having functional host system, verification, and FPGA framework layers.}

\begin{table}
\centering
    \begin{tabular}{p{3.1in}}
    
    \textbf{Core Layer} \\ \hline
    WISCV provides multiple core designs for pedagogical use \\
    Completely written in Verilog\\
    Reference designs, and skeleton files with interfaces for students for 3 designs (Single cycle, 5-stage pipeline, With cache)\\ \hline
    \textbf{Host System}\\ \hline
    RISC-V GNU Compiler Toolchain and Modelsim for verification and synthesis\\
    Verification and synthesis run on the command line via Makefile \\
     \hline
    \textbf{Verification} \\ \hline
    Full testbench for verifcation and trace generation\\
    Options for waveform visualization and DUT/reference per instruction register file comparison \\
    100s of pre-generated test programs \& complex test generator\\
    Can run in 3 modes: pipelined core (with or without cache), single-cycle core, or no core (to generate reference output)\\
    C implementation of LENET neural network\\ \hline

    \textbf{FPGA}\\ \hline
    Push-button FPGA synthesis script with support for cheap Xilinx devices (e.g. Arty A7 FPGA board) \\
On-the-fly loading of new programs, avoiding FPGA recompilation. \\ \hline

    \end{tabular}            
    \caption{Overview of WISCV Features}\label{tab:wiscv-features}
    \vspace{-0.3in}
\end{table}

\section{Our WISCV Implementation}\label{sec:wiscv}
Based on these insights, our experience, and surveying the course offerings and pedagogical goals of the computer architecture curriculum, we have developed the WISCV ``platform'' - for anonymized review it expands to Well Intentioned \& Systematic Computer in Verilog. It includes a complete 5-stage pipeline reference Verilog implementation (that instructors can use), for students a shell structure with interfaces defined and empty modules provided, two types of memories, modules to help build a cache, test generators, an extensive suite of generated tests for Verilog, a full processor test bench with checker scripts, and a wrapper around a push button FPGA flow compatible with inexpensive small FPGA boards from Xilinx. The entire repo with documentation is available at \url{https://anonymous.4open.science/r/wiscv-86E1}. Since it uses the RISC-V ISA, we are able to use the compiler and provide C tests and assembly tests. A list of WISCV features are in Table~\ref{tab:wiscv-features}. Our README on the anonymized github includes mode documentation on features and usage modalities. In particular, our FPGA infrastructure, allows new programs to be loaded into the FPGA's memory (in milliseconds), avoiding the need for FPGA recompilation to run different programs (20 minutes to an hour). The latter is a common flow in FPGA related courses.

This implementation achieves all of our pedagogical goals and goes further to give students a sense of ``accomplishment'' and satisfaction beyond toy assembly programs in a home-brew ISA. We also provide one simple deep-learning application - the classic LENET digit recognition, with quantized weights and implemented as a C program. In Verilog, our reference cores runs the application in simulation in 25 minutes. On FPGA, it runs in milliseconds. Our hypothesis was that this closes the loop nicely in exposure of topics to students and has a certain amount of wow factor - their core runs a cool AI application!

We trialled this infrastructure in a standard course offering where some students used the home-brew ISA, while a few groups were provided the option of using this infrastructure (which had some teething problems because of it's first deployment). Six students used our WISCV implementation, and we interviewed them individually at the end of the course for feedback (post assignment of grade). All team members got an A for the project portion. One of the authors was a TA for the course and hence had a window of opportunity to look at everyone's concerns. Summary of the interviews with annotated quotations from the students is in Table~\ref{tab:interviews}.

\begin{table}
\centering
    \begin{tabular}{p{3.1in}}
    \textbf{ISA Layer} \\ \hline
    Good exposure to real ISA: (1) 
    ``Many helpful online resources to facilitate understanding''. (2)
    ``Simplified datapath in comparison to internal ISA'' \\ \hline
    \textbf{Testing} \\ \hline
    Easier to understand C programs vs. complex assembly tests. 
     ``Presence of C tests allows us to understand tests easily'' \\
     \hline

    \textbf{LENET demo} \\ \hline
    Having a cool demo is a great incentive to work on full integration. (1)
    ``Provides bragging rights '' (2)
    ``can mention this to interviewers'' \\ \hline

    \textbf{FPGA} \\ \hline
    Having a FPGA run your CPU gives sense of more ownership of the project. (1)
    ``FPGA is a good incentive for students; feels more special and exciting''. (2)
    ``Running the project on FPGA is good for teaching about hardware limitations'' \\ \hline
    \end{tabular}            
    \caption{Student Assessment of WISCV with summary and quotes from our transcribed notes.}\label{tab:interviews}
    \vspace{-0.3in}
\end{table}

\subsection{Uses by other instructors and Ecosystems}
While none of the RISC-V implementations we analyzed perfectly instantiated the full ecosystem, we believe WISCV's purposeful design and implementation allow turn-key deployment of all components or different disaggragated pieces. We enumerate a few below. 
\textit{Microarchitecture lite}: Here an instructor could do a simple two-stage or single-cycle processor only, but still retain all of the other pieces. They can use our {\tt nopipe} mode in the code release.
\textit{FPGA-lite}: Due to time or resource limitations, the FPGA layer can be completely eliminated with all the other pieces being used as is.
\textit{Verification lite:} We provided nearly 250 tests and an extensive test generator. Students getting their implementation to be completely correct is a bit time consuming and may not fall within the pedagogical goals. Instructors can pick and choose which tests to use (e.g. just use unit tests).
\textit{Performance heavy:} Building on top of verification lite is another option, where instructors could focus on students writing C or assembly programs that teased apart performance issues, with the goal being students learning how to write such microbenchmarking code. High impact versions of such research is common ~\cite{abdelkhalik2022demystifying,sun2022dissecting}, and thus such training is valuable.

\section{Conclusion}
% With the recent passage of the CHIPS and Science Act in the United States, \$52.7B of federal funds have been allocated for investment in semiconductor research, development and manufacturing. Along side this, the explosion of Deep Learning has sparked new interest in chip design for accelerating these compute hungry workloads with many chips being built: most cloud-service providers are building their own chips, as are companies like Tesla and GM/Cruise for their vertical uses, and over 100 startups.

With this explosive (re-)emerging interest in computer architecture and semiconductor technologies due to demand for Deep Learning accelerators and the recent passage of the CHIPS and Science Act, institutions will likely want to give thought to how they might restructure their introductory digital design courses to keep pace with this new era of semiconductor research and development. This paper underscores the importance of the ecosystem which accompanies a given architecture's implementation, and examines the state of the ecosystems across many popular RISC-V implementations for the purpose of elucidating what implementations should be considered given a particular pedagogical use-case. We have developed and released, for public use, a comprehensive pedagogical framework, that we are deploying and plan to continue supporting. The student feedback validates many of our hypotheses on the benefits of moving to a production ISA.

\bibliographystyle{ACM-Reference-Format}
\bibliography{sample-base}

\end{document}